\documentclass[final]{elsarticle}
\usepackage{hyperref}
\usepackage{graphicx}
\usepackage{subcaption}
\usepackage{bm}
\usepackage{amssymb}
\usepackage{natbib}
\begin{document}

\begin{frontmatter}
\title{Career choice as an extended  \\ spatial evolutionary public goods game}

\author{Yuan Cheng\corref{cor1}}
\ead{\{yuancheng, yxue, mengchang\}@careersciencelab.com}
\author{Yanbo Xue\corref{}}
\author{Meng Chang\corref{}}
\address{Career Science Lab,  Beijing, China}
\cortext[cor1]{Corresponding author}

\begin{abstract}
We propose an extended spatial evolutionary public goods game (SEPGG) model to study the dynamics of individual career choice and the corresponding social output. Based on the social value orientation theory, we categorized two classes of work, namely the public work if it serves public interests, and the private work if it serves personal interests. Under the context of SEPGG, choosing public work is to cooperate and choosing private work is to defect. We then investigate the effects of employee productivity, human capital and external subsidies on individual career choices of the two work types, as well as the overall social welfare. From simulation results, we found that when employee productivity of public work is low, people are more willing to enter the private sector. Although this will make both the effort level and human capital of individuals doing private work higher than those engaging in public work, the total outcome of the private sector is still lower than that of the public sector provided a low level of public subsidies. When the employee productivity is higher for public work, a certain amount of subsidy can greatly improve system output. On the contrary, when the employee productivity of public work is low, provisions of subsidy to the public sector can result in a decline in social output. 

\end{abstract}
\end{frontmatter}

\newpage
\section{Introduction}\label{section:intro}
Career choice is an important research topic that has drawn attentions mostly from the psychology community. The most widely used models consider three factors in the decision-making process of a career choice: {\em The intrinsic factor} of personal interests and job satisfaction, {\em the extrinsic factor} of job availability and scope of occupation, and {\em the interpersonal factor} of influences from friends, family and society \cite{carpenter1977career, beynon1998visible}. The complexity of a career-choosing process is reflected by the interplay of two mechanisms, one that justifies the suitability of a person to the target career based on working skills, abilities, characteristics, and interests, and one that is driven by external environment conditions \cite{career2010}. To this end, career choice study has been divided into two camps. The first camp focuses on individual's internal factors such as needs, motivations, values, abilities and personality, whose principles and research methods are well known in psychology \cite{aycan2003, rojewski2003, arthur2005, jiang2018}. The second camp marries psychology to statistics and data science to equate career choice with person-job matching problem, and the individual psychological differences are only included as subjective and objective conditions  \cite{chuang2016,bednarska2017,van2018}. Without belittling the progresses made so far, over the last few decades, we have witnessed an unfolding pattern of career choice study to include more and more subtle dimensions (see \cite{akosah2018systematic} and the references therein). 

In parallel to the study of career choice, researchers in social psychology have made much progresses in understanding the common natures across societies and cultures, which leads to the birth of  {\em The Theory of Basic Human Values}  \cite{schwartz1992universals}.  In The Theory of Basic Human Values, ten universal values organized into four groups: openness to change, self enhancement, conservation and self-transcendence. In 2012, Schwartz expanded the basic individual values into a set of 19 elements \cite{schwartz2012refining}. Undoubtedly, values play an essential role in a person's career choice. Many empirical studies have suggested that one is more likely to choose a career that matches his/her personal values \cite{de2008value, brown2002role}. The study of human values has led us to a modular way of looking at career choice. If data-centric method is the key to resolving the job availability and scope problem, career choice can be simplified as a social value orientation (SVO) problem \cite{forsyth2006conflict}. In SVO, people can be classified into two groups, the {\em proself/egoistic} group and the {\em prosocial/altruistic} group. The proself person maximizes their own gains and the prosocial person help others by sharing their gains with the society. 

Now that the driving force for a career choice has been attributed to a person's value orientation, it is natural to study the pathway of the formation of human values. Much like a successful career choice will bestow a person benefits like promotion, salary increase and social achievement, an unsuccessful work transition will discourage a person in making similar changes in the future. This is not only true to the experiencing person, but to the witnesses. This phenomenon can be summarized by the social learning theory (SLT) due to Bandura's orignal work \cite{bandura1963social} and extensions  \cite{krumboltz1976}. In SLT, learning requires social context, information flow, cognition, behaviour changes and reinforcement. The most prominent feature is that people imitates others' behaviours by modelling its performance in a predictive fashion. In the context of career choice, SLT offers a framework to modelling the propagation of human values. For example, many people are interested in serving the public to increase others' well-being. But working in the public service sector and donating to a charity are two different ways of helping others \cite{dur2018}. Studies show that different social context yields different altruistic behaviours, possibly explained by SLT. 

Other than the socio-psychological way of studying the impacts of human values on career choice, we can quantitatively study the dynamics of the career changing behaviours. A complex physical system can better characterize the interactions between individuals or groups as a Public Goods Game (PGG)
 \cite{perc2013}.  In a PGG, public goods such as social benefit systems or the environment are particularly prone to be exploited  by individuals at the expense of others.  Analogous to the famous prisoner’s dilemma that focuses on the human pairwise interactions, PGG studies behaviours involving multiple players. Among the players, cooperators contribute a cost to the public pool, while defectors contribute nothing. The total contribution is multiplied by a factor reflecting a synergetic effects of cooperation. Then the resulting reward is equally redistributed to all players, regardless their strategies.  Experiment shows that, for a well-mixed population, defectors will dominate the whole game if the synergy factor is less than the total number of players \cite{hauert2006synergy}. The collective free-riding behaviours of defectors would result in ``tragedy of the commons'' \cite{hardin1968}. 
 
Recently, there are two practical extensions to PGG. Firstly, spatial structures have been considered in the study of game strategies, {\em i.e.}, lattice or densely connected networks \cite{kim2015spatial}. Secondly, the evolution of game dynamics has been included to study the critical points or phase transition of PGG \cite{lahkar2019evolutionary}. Spatial evolutionary PGG (SEPGG) has become a powerful framework to study games \cite{nowak1992, szabo1998}. Not only fruitful theoretical results have been published \cite{szabo2007, roca2009,perc2010,perc2017}, but many attempts have studied issues like social dilemma \cite{kollock1998}. Not surprisingly, factors like punishment \cite{szolnoki2011, helbing2010}, diversity \cite{shi2010,zhu2014, huang2015}, volunteering \cite{hauert2002, semmann2003},and  reputation mechanism \cite{milinski2002, xia2016} have been found to play an important role in promoting cooperative behaviours in SEPGG. However, most existing researches are still based on toy model like two-player games and they are rarely applicable to real-world problems.

In this paper, we inherit the SEPGG model to study the career choice problem. To reflect on the intrinsic differences among job candidates, the job output is characterized by a multiplication of human capital, employee productivity and personal effort. For the sake of simplicity and without loss of generality, we narrow down the career choices into value orientations of two distinct types of works, the one serving public interests ({\em public work}) and the one serving personal interests ({\em private work}). In reference to the PGG model, choosing a public work is to cooperate and choosing a private work is to defect. A further extension is also made to include subsidy effect into our model. Due to the agent-level modelling capability, our extended SEPGG lays down the foundation to study stable patterns and emergent behaviours of a career choice process from a complex system perspective \cite{anzola2017}. To be specific, our computational experiments study the influences of employee productivity and external subsidy on the welfare of the overall system.  Our best effort of literature search shows that this work is the first attempt to study the career choice problem using an extended SEPGG model. 

The remainder of the paper is organized as follows. Section \ref{section:model} presents our extended SEPGG model and the game dynamics. In Section \ref{section:simulation}, numerical simulations are conducted thoroughly and the simulation results are analyzed. Finally, conclusions are provided in Section \ref{section:conclusion}. 

\section{Model} \label{section:model}
As mentioned in Section \ref{section:intro}, our model assumes two distinct types of career choice: the {\em public work}, whose values are serving public interests, and the {\em private work}, whose values are serving personal interests. The two types are denoted by $P$ and $R$ respectively. For a type-$P$ work, a person contributes to the common pool and every player in the game benefits. For a type-$R$ work, a person only contributes to his/her own welfare and no other player benefits. In our model, a person $i$ can choose between a public work ($s_i=P$) and a private work($s_i=R$).

\subsection{Employee productivity, human capital and work effort}
To decide upon how much a worker can produce, next we will introduce three factors, namely the employee productivity, the human capital and the work effort, elaborated as follows:
\begin{itemize}
	\item {\bf Employee productivity}:  Employee productivity is represented by $r_R$ for private work and $r_P$ for public work, respectively. The value of employee productivity provides a measure of work efficiency at an individual level. Given other factors fixed, a higher  employee productivity implies a greater output of the work, no matter it is type-$P$ or type-$R$.
	\item {\bf Human capital}: Human capital aggregates many personal attributes of a person, including knowledge, skills, personality, and creativity, into an economic metric. A higher human capital generally promises a higher output of individual labour.  The human capital of person $j$ is denoted as $\gamma_j \in [0,1]$. Although human capital can be improved in one's career, mainly through education and training, it is assumed to be unchanged in each instance of our experiment.	
	\item {\bf Work effort}:  In addition to choosing the type of work, a person can also decide on the level of effort, expressed by $e_j \in [0,1]$ for person $j$. When $e_j=0$, it means that the person is accepted by the job, however, he/she has not put in any effort. On the contrary, when $e_j=1$, it means that the person will devote his/her best effort to the job. 
\end{itemize}	

\subsection{Work output, utility, and subsidy effect}
For the $j$-th person making career choice of a type-$P$ work, his/her work output  is the multiplication of human capital $\gamma_j$, employee productivity $r_P$, and work effort $e_j$, expressed by $\gamma_j r_P e_j$. The altruistic nature of type-$P$ work makes this person a cooperator in the SEPGG. Let us define $N_j$ as the {\em closed neighbourhood} of person $j$,  {\em i.e.}, $N_j \triangleq N(j) \cup j $.  Following the output allocation strategy in an SEPGG, this output will be equally shared among all neighbouring person $N(j)$, in addition to the $j$-th person.  Therefore,  for any person $i \in N_j$, his/her utility due to the work output of person $j$ can be expressed as
\begin{equation}
    u_{i\leftarrow j} = \frac{\gamma_j r_P e_j}{|N_j|} ,   \forall i \in N_j, \label{eqn:utility_p}
	\label{eqn1}
\end{equation}
where $|N_j| $ denotes the size of person $j$'s closed neighbourhood.

It is noteworthy to mention that $i=j$ is not excluded in Eqn. (\ref{eqn:utility_p}), which means that person $j$ also benefits from his/her own work output, equally as other neighbours. 

The output for a type-$R$ work follows a similar definition, given as $\gamma_{j} r_R e_{j}$ for the $j$-th person. The only difference is that the employee productivity has been replaced by $r_R$. Since it is the nature of a type-$R$ work to {\em defect} in the SEPGG, the utility function of  a person $i \in N_j$ will be mostly zero other than that $i=j$, that is
\begin{equation}
u_{i \leftarrow j} = \left\{
 \begin{array}{rcl}
    \gamma_{j} r_R e_{j},   &\textrm{if $i = j$} \\
    0,                    &\textrm{otherwise}.
 \end{array}
\right.\label{eqn:utility_r}
\label{eqn2}
\end{equation}

Having defined the utility function for all person in our game, the total return of the $i$-th person in the SEPGG can be calculated by the following summation,
\begin{equation}
	o_i = \sum_{j\in N_i} u_{i \leftarrow j},
	\label{eqn3}
\end{equation}
where the utility function for $i\in N_j$ is chosen from either Eqn. (\ref{eqn:utility_p}) or Eqn. (\ref{eqn:utility_r}) depending on person $i$'s career-choosing strategy.

Additionally, we add subsidy effect to our model to reflect on incentives from external environment toward the type-$P$ work. The total return for person $i$ is as follows,
\begin{equation}
u_i=\left\{
 \begin{array}{rcl}
    o_i + h,   &\textrm{if $s_i = P$} \\
    o_i,       &\textrm{if $s_i = R$}
 \end{array}
\right.
\label{eqn4}
\end{equation}
where $h$ is the external subsidy to incentivize people to serve the public. Likewise, it can also be understood as a utility value obtained by the altruistic preferences of individuals engaged in public service work \cite{dur2018}. 

Finally, the net output of the entire system (excluding subsidy) is as follows,
\begin{equation}
	o = \sum_{i\in \Omega} u_{i \leftarrow j}
	\label{eqn5}
\end{equation}
where $\Omega$ represents the set of all players in the game. The system output can be considered as the social welfare.

\subsection{Game dynamics}
In our extended SEPGG model, each person plays a game with his/her neighbours using a strategy similar to the standard PGG. We adopt the spatial structure of a two-dimensional square lattice with periodic boundary conditions. Each site on the lattice is occupied by one person with four neighbours (von Neumann neighbourhood). Accordingly, each person belongs to maximally five different groups. For the $i$-th person, his/her strategy is executed by controlling two factors: the work type $s_i \in\{R,P\}$ and the level of work effort $e_i \in [0,1]$.

In designing the game dynamics, we employ the core ideas of social learning theory, that is, a person learns through observations of others' behaviours and chooses career that reflect the common social values  \cite{krumboltz1976, sigmund2010}. Mathematically, we use Fermi's rule to update a person's career choice  \cite{szabo2002}. When the $i$-th person needs to revise his/her strategy, a randomly chosen neighbour $j$ is imitated. Given the total returns $u_i$ and $u_j$ of person $i$ and person $j$ at step $t$ according to Eqn. (\ref{eqn4}),  the probability of person $i$ adopting person $j$'s strategy at step $t+1$  satisfies the following Fermi-Dirac (F-D) distribution:
\begin{equation}
	w_{i \leftarrow j} = \frac{1}{1 + \exp\{(u_{i} - u_{j} + \tau)/k\}},
	\label{eqn6}
\end{equation}
where $\tau \geq 0$ is the cost of work transition and $k$ denotes the irrational degree of individual decision-making. In the F-D distribution, $k$ can be considered as a temperature effect. A higher $k$ makes a person irrational in his/her decision-making. When $k>0$, under-performing strategies may also be selected due to individual mistakes in the observation and evaluation of the utilities. When $k \rightarrow 0$, a person will deterministically  follow another person's strategy.

Next, we discuss two extensions of the game dynamics:
\begin{itemize}
	\item {\bf Work type adoption}: When person $i$'s work type is different from that of person $j$, {\em i.e.}, $s_i(t) \neq s_j(t)$, the work effort of person $i$ remains unchanged from step $t$ to $t+1$, {\em i.e.}, $e_i (t+1)= e_i(t)$. Under such circumstances, $s_i$ will be converted to $s_j$ with probability $w_{i \leftarrow j}$ at $t+1$, expressed as, 
\begin{equation}
s_{i}(t+1) = \left\{
 \begin{array}{rcl}
    s_j(t),   &\textrm{with prob. $w_{i \leftarrow j}$ in Eqn. (\ref{eqn6})} \\
    s_i(t),   &\textrm{otherwise}.
 \end{array}
\right.
\label{eqn7}
\end{equation}

	\item {\bf Work effort adoption}: When person $i$  and person $j$ have the same work type at step $t$, {\em i.e.}, $s_i(t) = s_j(t)$, there is no need for person $i$ to change work type at step $t+1$, {\em i.e.}, $s_i (t+1)=s_i (t)$.  In imitating person $j$'s strategy, the work effort of person $i$ will consequently increase or decrease depending on if $e_i(t)$ is greater or smaller than that of person $j$.  The work effort adoption dynamics can be characterized as follows,
\begin{equation}
e_{i}(t+1) = \left\{
 \begin{array}{ccl}
    m(t+1),   &\textrm{with prob. $w_{i \leftarrow j}$ in Eqn. (\ref{eqn6})} \\
    e_i(t),   &\textrm{otherwise},
 \end{array}
\right.
\label{eqn8}
\end{equation}
where $m(t+1) \sim U(e_i (t),1]$ if $e_j (t) > e_i (t)$ and $m(t+1) \sim U[0, e_i (t))$ if $e_j(t) < e_i (t)$.  Here, $U[\cdot, \cdot]$ represents a uniform distribution. 
\end{itemize}

Now, we have extended the standard SEPGG to include more realistic dynamics, exemplified by a career-choice game. For the special case of $r_R=1$, $\gamma_i = 1$ and $e_i =1$, our extended SEPGG will reduce to a standard SEPGG model as follows
\begin{equation}
u_{i} = \left\{
 \begin{array}{ccl}
    r_P n_R/(n_P + n_R),       &\textrm{if $s_i = P$} \\
    r_P n_R/(n_P + n_R) + 1,   &\textrm{if $s_i = R$},
 \end{array}
\right.
\label{eqn9}
\end{equation}
where $n_R$ and $n_P$ are the number of people who choose type-$P$ and type-$R$ work in person $i$'s closed neighbourhood, {\em e.g.}, $n_P+n_R=5$ in our model. The standard SEPGG has been widely studied as a social dilemma for group interactions. When $r_P$ is relatively small ($r_P<5$), it is clear that the benefit of type-$R$ work is always higher than that of  type-$P$ work.Therefore, choosing type-$R$ work is a dominant strategy. The Nash equilibrium of the SEPGG will be all choose to defect.

\section{Simulation results} \label{section:simulation}
\begin{figure}[!htb]
     \centering
     \includegraphics[width=0.80\textwidth]{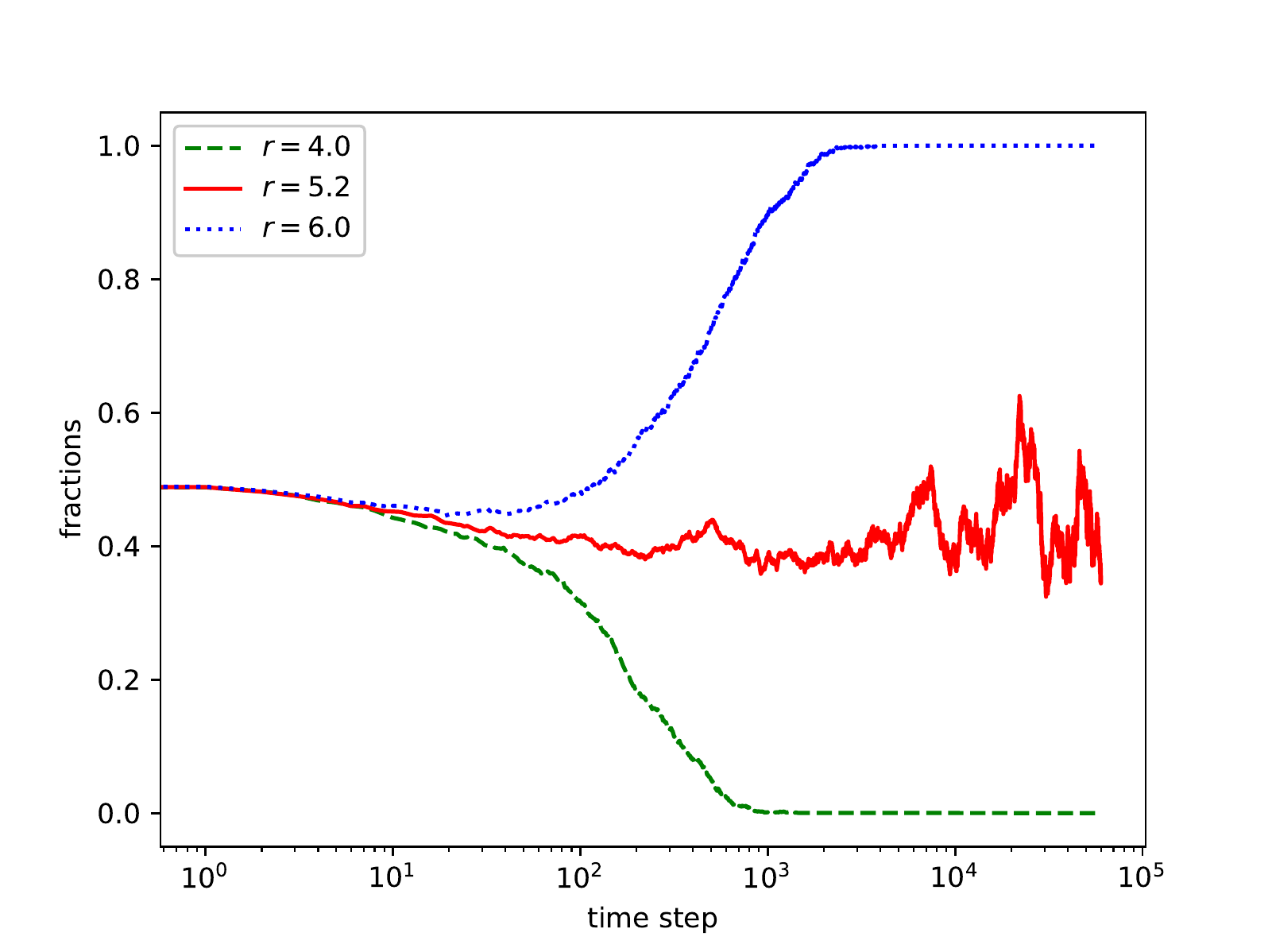}
     \caption{Fraction of type-$P$ work as a function of time step with $h = 0$ and $r_P = 4, 5.2, 6.0$.}
     \label{fig:fig1}
\end{figure}
Our simulation is conducted based on the extended SEPGG model described in Section \ref{section:model}. The spatial structure of the game is a square lattice of $100$ by $100$. We set $k = 0.1$ and $\tau = 0$ by assuming no additional cost in the career transition process \cite{szabo2002}.  To initialize our game, type-$P$ work and type-$R$ work are uniformly distributed on the lattice sites. The initial values of human capital and work effort for each person are drawn from a uniform distribution $U[0,1]$. A Monte-Carlo simulation of $5,000$ instances are conducted and the results are averaged. To allow for recovery of the system from the {\em initialization shock}, we use a washout time of $55,000$ steps. When we update the strategies for all participants at any specific time, we use an asynchronous method, which sequentially chooses a random person to update according to Eqns. (\ref{eqn7}) and (\ref{eqn8}) till all strategies are updated. For simplicity, we assume that the employee productivity of type-$R$ work to be 1 throughout our experiments, {\em i.e.}, $r_R = 1$.

\begin{figure}[htb]
    \centering
    \includegraphics[width=\textwidth]{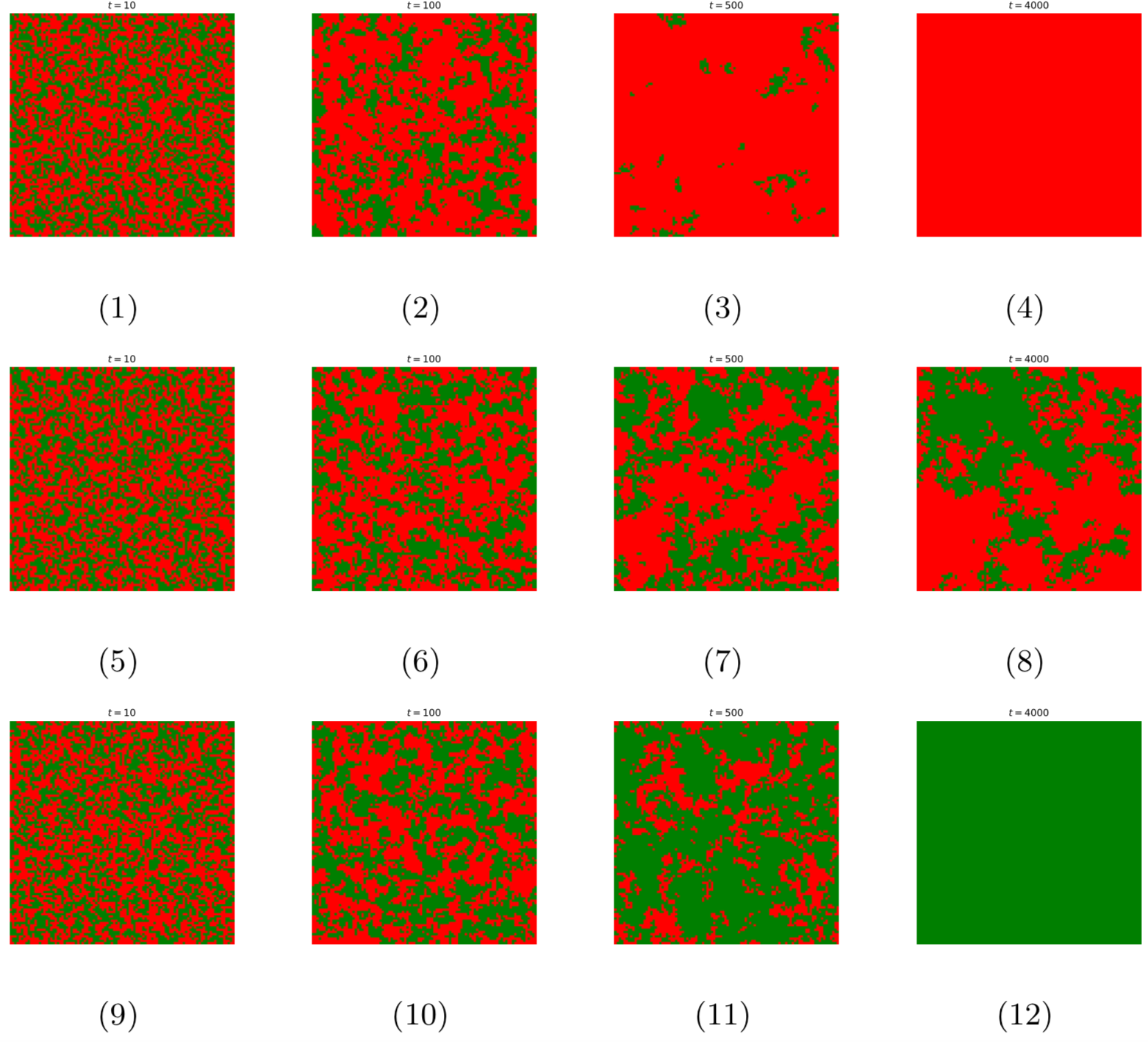}
    \caption{Snapshots of SEPGG with $h = 0$ and $r_P$ =4, 5.2 and 6 at $t$ = 10, 100, 500 and 4000. Green color and red color indicates type-$P$ and type-$R$ work respectively.}
    \label{fig:fig2}
\end{figure}

\subsection*{(1) Three outcomes of a game: all cooperate, all defect, or a mixture}
In our model, there are only three possible outcomes: all participants choose type-$P$ work, all participants choose type-$R$ work, or a mixture of the two. Figure \ref{fig:fig1} demonstrates all three cases by plotting the fraction of participants who chooses type-$P$ work as time evolves. It can be seen that, when simulation reaches $10,000$ steps, most games approach a stable state. The varying parameter is the employee productivity of type-$P$ work, $r_P$. Specifically speaking, for $r_P = 4$, our model converges to a 100\% of type-$R$ work, and for $r_P = 6$, every player in our game chooses a type-$P$ work. For the case of $r_P = 5.2$, the percentage of type-$P$ work fluctuates between 37\% and 63\%.

To further speculate on the game dynamics, we take snapshots of the game at different time steps under the same parameter settings of $r_P$.  As seen in Figures \ref{fig:fig2}(1) - \ref{fig:fig2}(4), when $r_P = 4$, more and more participants escape from type-$P$ work. For $r_P = 6$, the exact opposite pattern can also be observed from Figures \ref{fig:fig2}(9) - \ref{fig:fig2}(12), where more and more participants choose type-$P$ work till every player in the game has the same work type. When $r_P = 5.2$, a critical point has been discovered, as shown by Figures  \ref{fig:fig2}(5) - \ref{fig:fig2}(8), that type-$P$ work and type-$R$ work are still highly mixed after  $t = 4,000$.

\subsection*{(2) Employee productivity and phase transition}
\begin{figure}[!htb]
     \centering
     \includegraphics[width=0.50\textwidth]{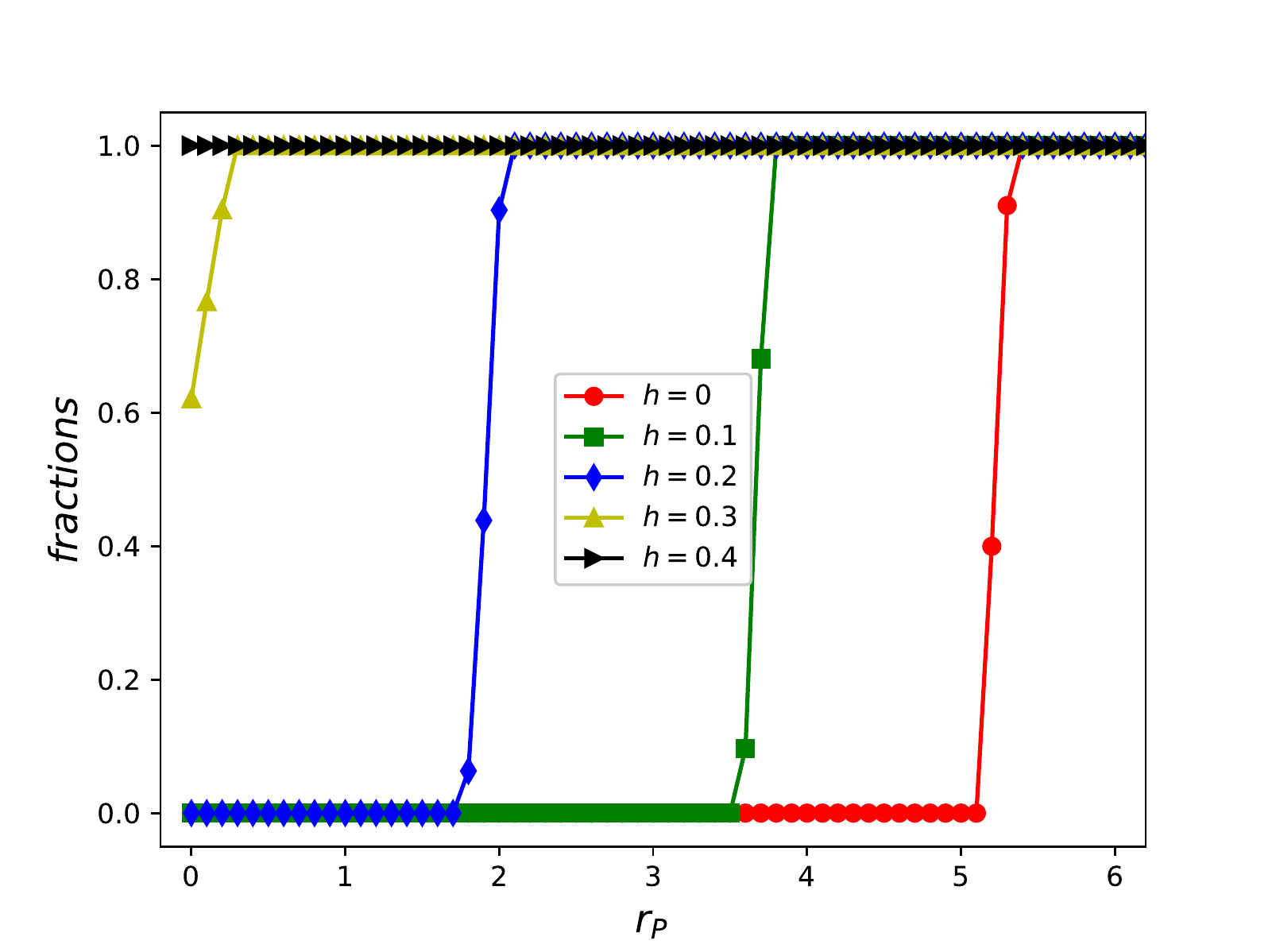}
     \caption{Fraction of public work as a function of $r_P$ with $h$ = 0, 0.1, 0.2, 0.3 and 0.4}
     \label{fig:fig3}
\end{figure}

For a thorough study, we first investigate the impact of type-$P$ employee productivity on the fraction of people who ends up choosing public work. The results for different settings of subsidy level are shown in Figure \ref{fig:fig3}. When subsidy $h = 0$, the system undergoes a phase transition at $r_P \in [5.1, 5.4]$. All individuals choose private work as $r_P<5.1$ while all individuals choose public work as $r_P > 5.4$. During the phase-transitioning stage, the system is in a mixed state of type-$P$ and type-$R$ works. Similar patterns can be observed with an increasing external subsidy, although phase transition happens at a reduced level of $r_P$. For example, when $h=0.1$, the phase transition occurs at $r_P \in [3.5,3.8]$, and when $h = 0.2$, it occurs at near $r_P \in [1.7, 2.0]$.  When the subsidy level reaches $h = 0.3$, there are 61\%  of type-$P$ work in the system despite of $r_P = 0$.  In the case of $h = 0.4$, regardless of the value of $r_P$, all participants choose public work.

The results herein show that a person will choose public work only after his/her productivity reaches a certain value. In addition, external subsidies incentivize less productive people to switch to type-$P$ work, as we expect. When the subsidy is large enough, {\em e.g.}, $h=0.4$, all participants choose the type-$P$ work, even if their productivity is much less than that of the type-$R$ work, that is, $r_P \ll r_R$.

\begin{figure}[!htb]
    \centering
    \begin{subfigure}[b]{0.43\textwidth}
        \includegraphics[width=\textwidth]{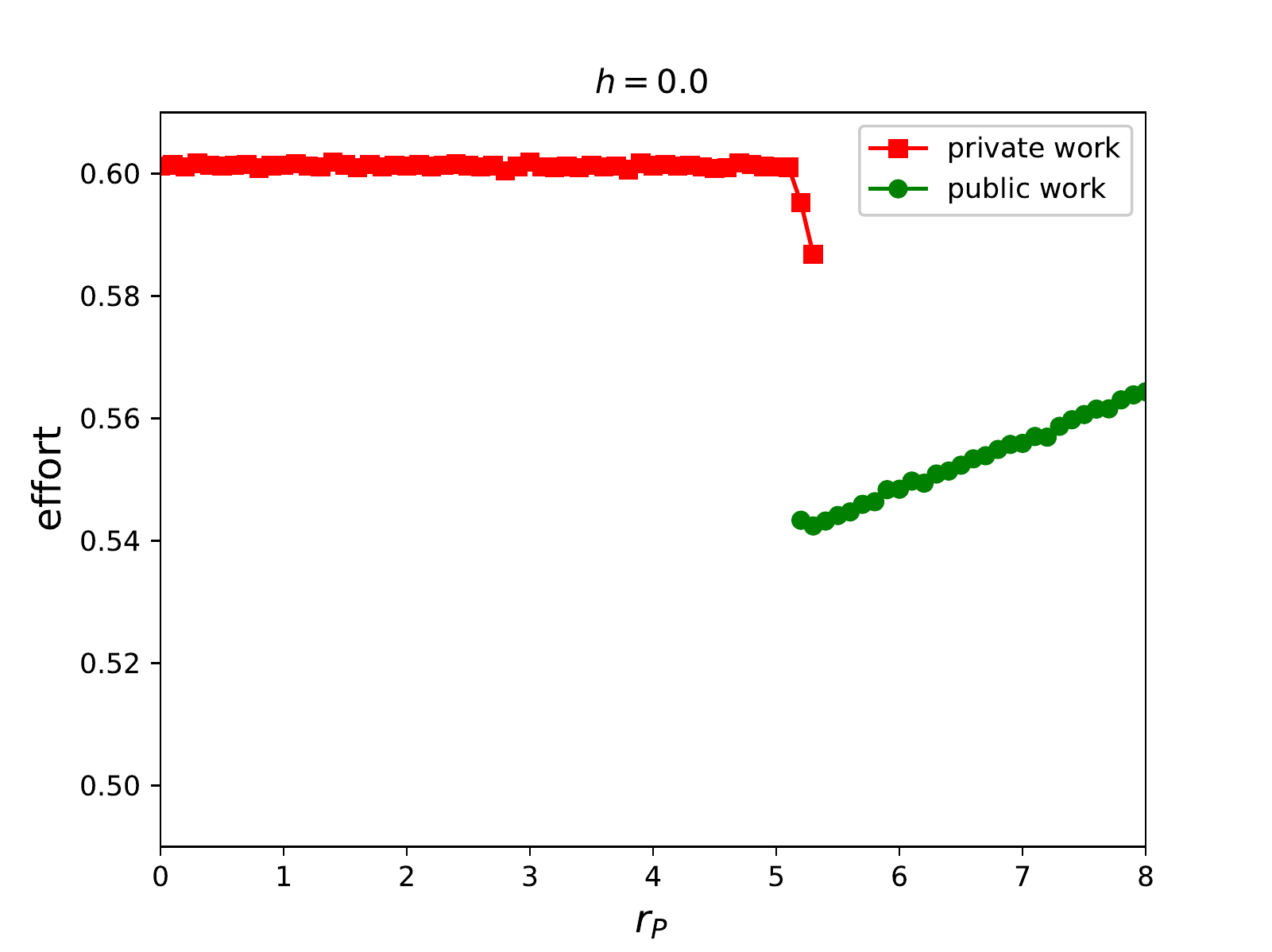}
        \caption{}
        \label{fig:fig4_1}
    \end{subfigure}
    \begin{subfigure}[b]{0.43\textwidth}
        \includegraphics[width=\textwidth]{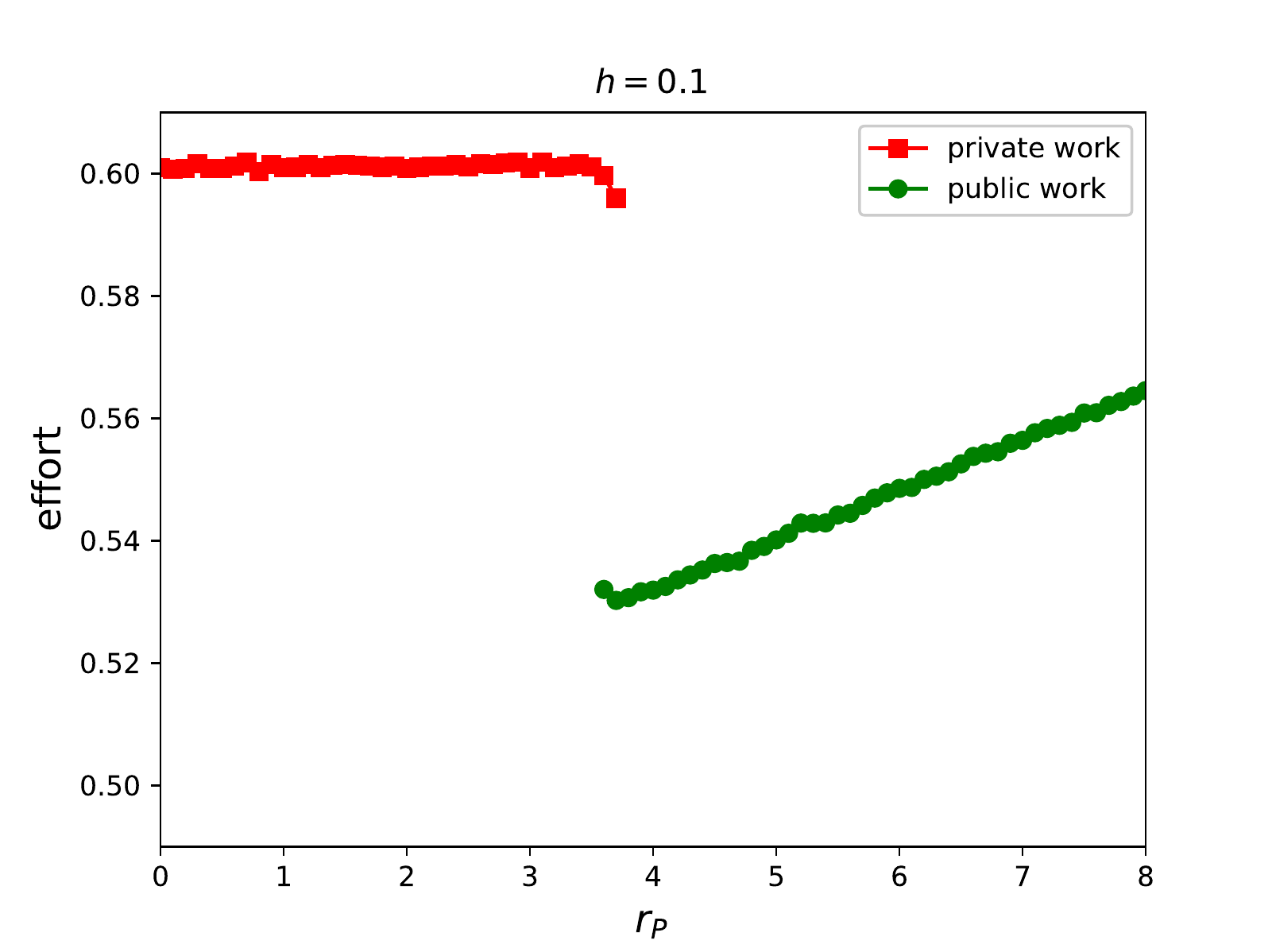}
        \caption{}
        \label{fig:fig4_2}
    \end{subfigure}
    \begin{subfigure}[b]{0.43\textwidth}
        \includegraphics[width=\textwidth]{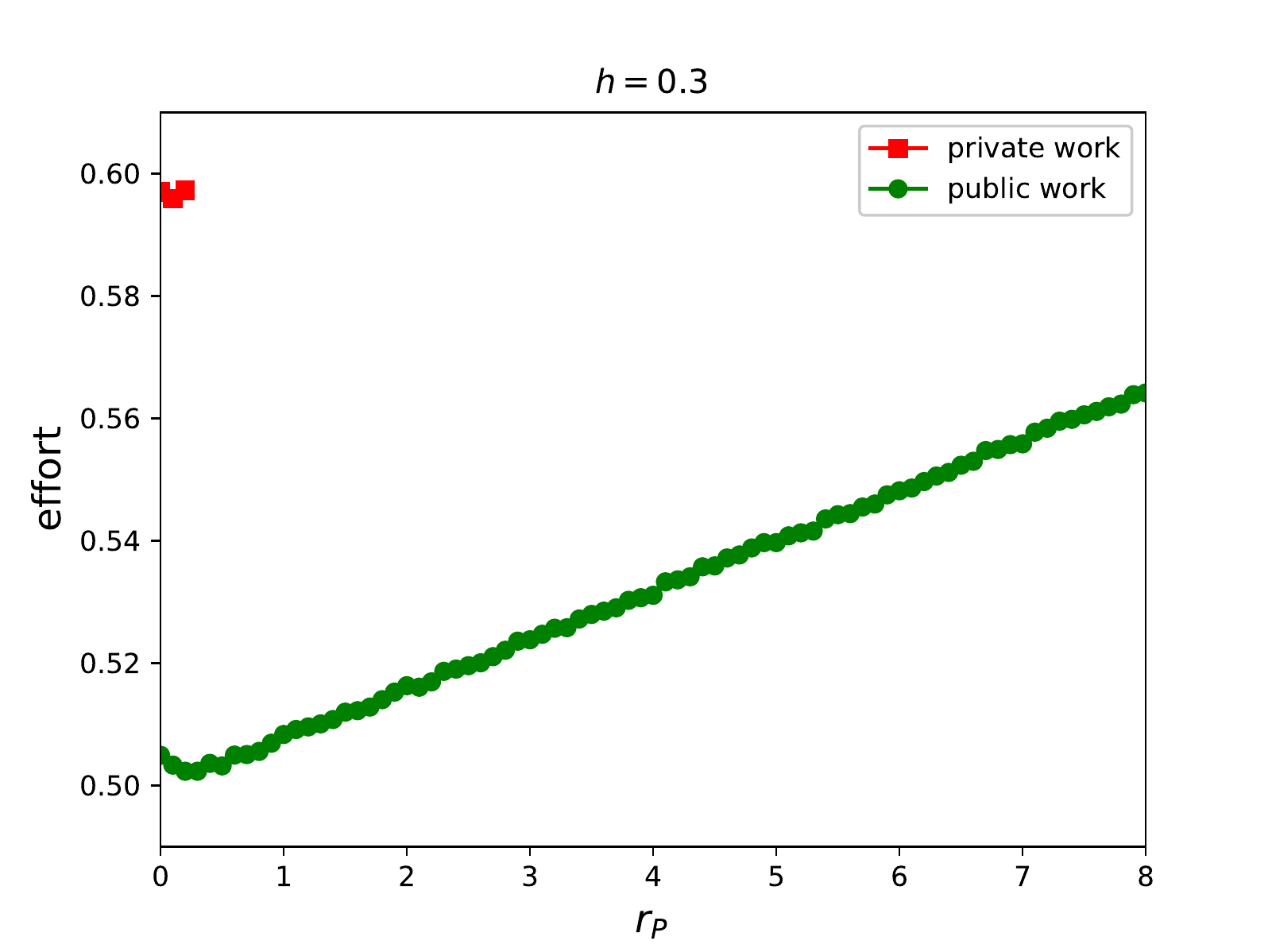}
        \caption{}
        \label{fig:fig4_3}
    \end{subfigure}
    \begin{subfigure}[b]{0.43\textwidth}
        \includegraphics[width=\textwidth]{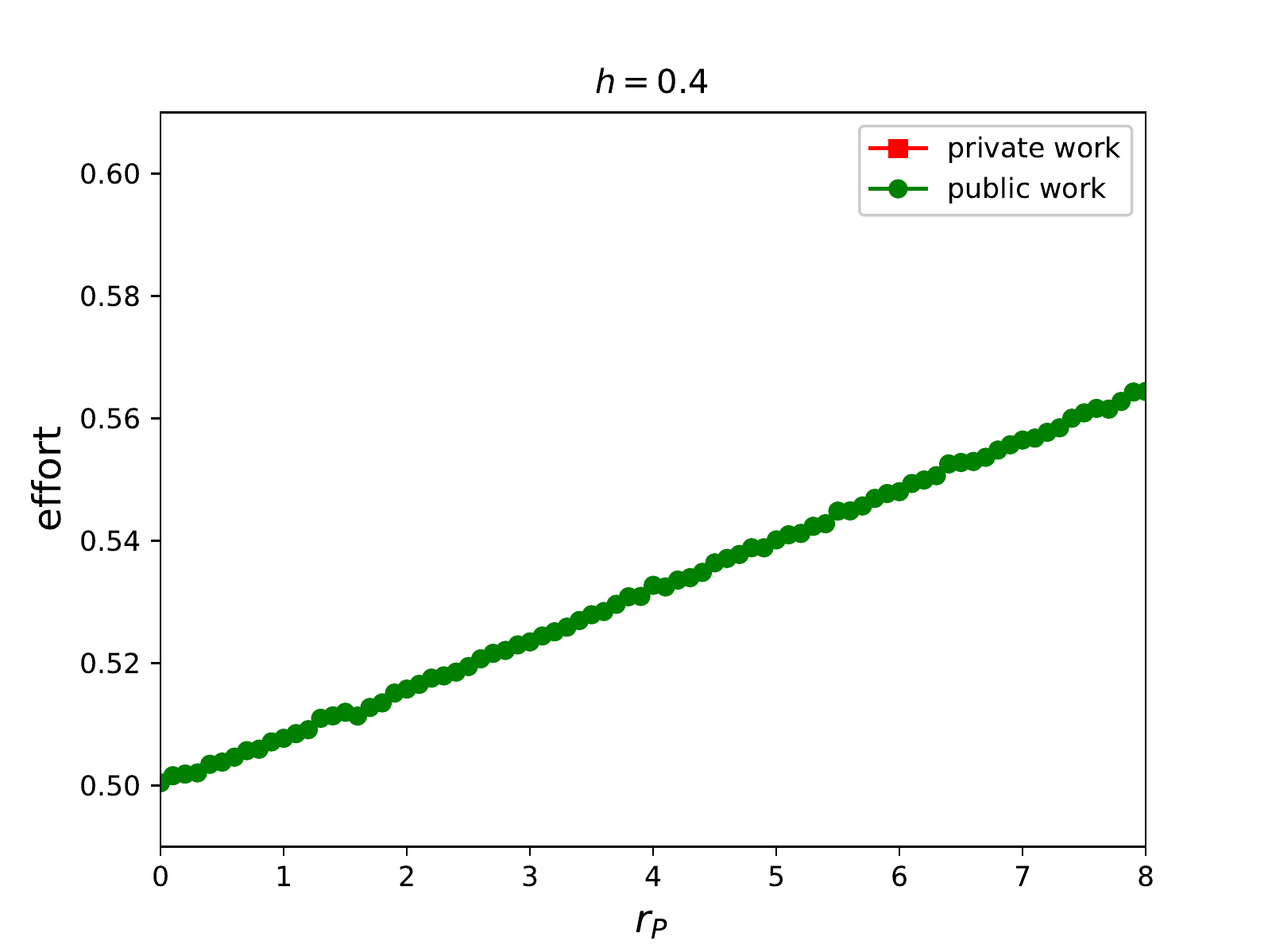}
        \caption{}
        \label{fig:fig4_4}
    \end{subfigure}
    \caption{Effort as a function of $r_P$ for public and private work with $h$ = 0, 0.1, 0.3 and 0.4.}
    \label{fig:fig4}
\end{figure}

\subsection*{(3) Work effort and human capital}

\begin{figure}[htb]
    \centering
    \begin{subfigure}[b]{0.43\textwidth}
        \includegraphics[width=\textwidth]{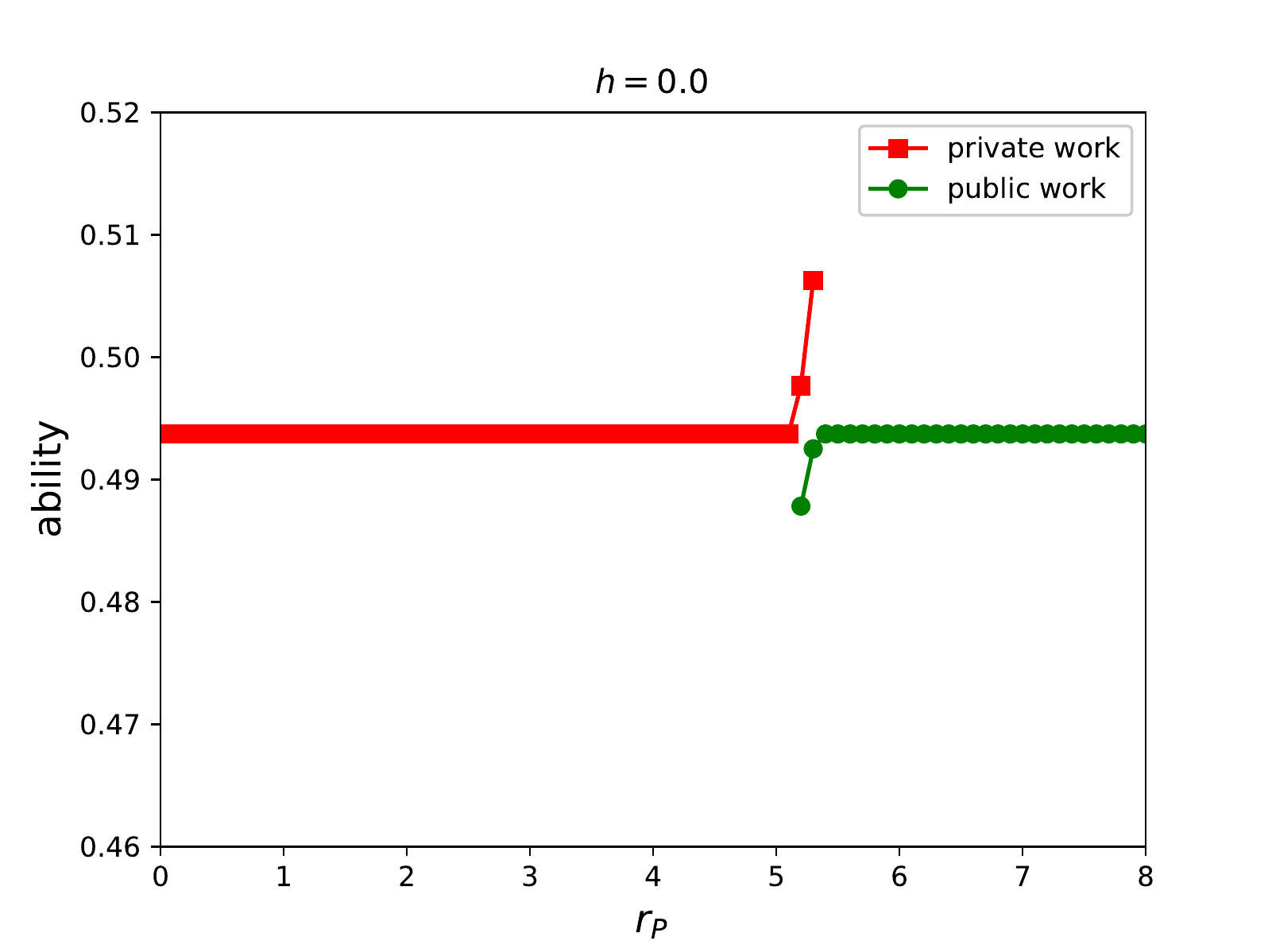}
        \caption{}
        \label{fig:fig5_1}
    \end{subfigure}
    \begin{subfigure}[b]{0.43\textwidth}
        \includegraphics[width=\textwidth]{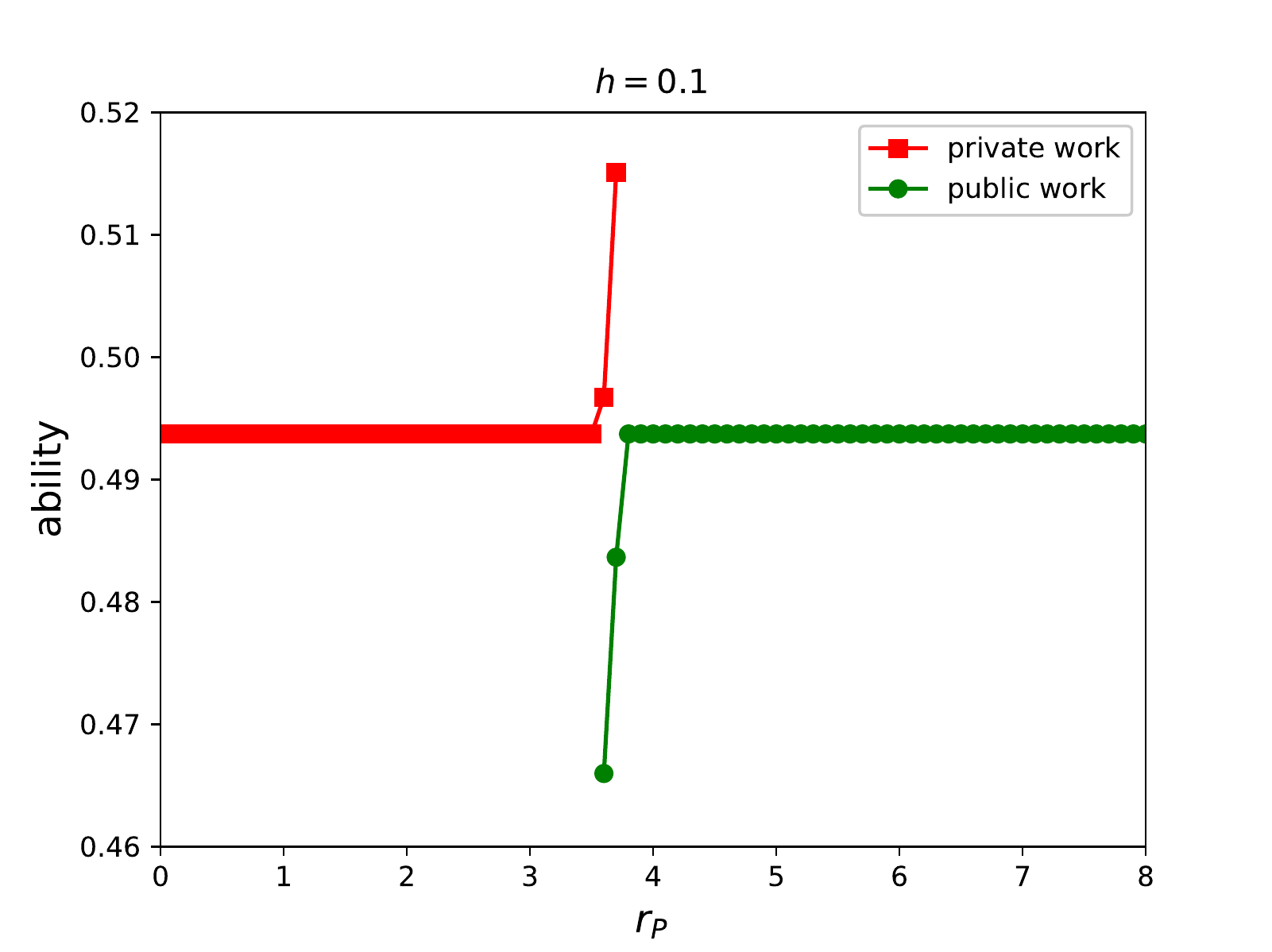}
        \caption{}
        \label{fig:fig5_2}
    \end{subfigure}
    \begin{subfigure}[b]{0.43\textwidth}
        \includegraphics[width=\textwidth]{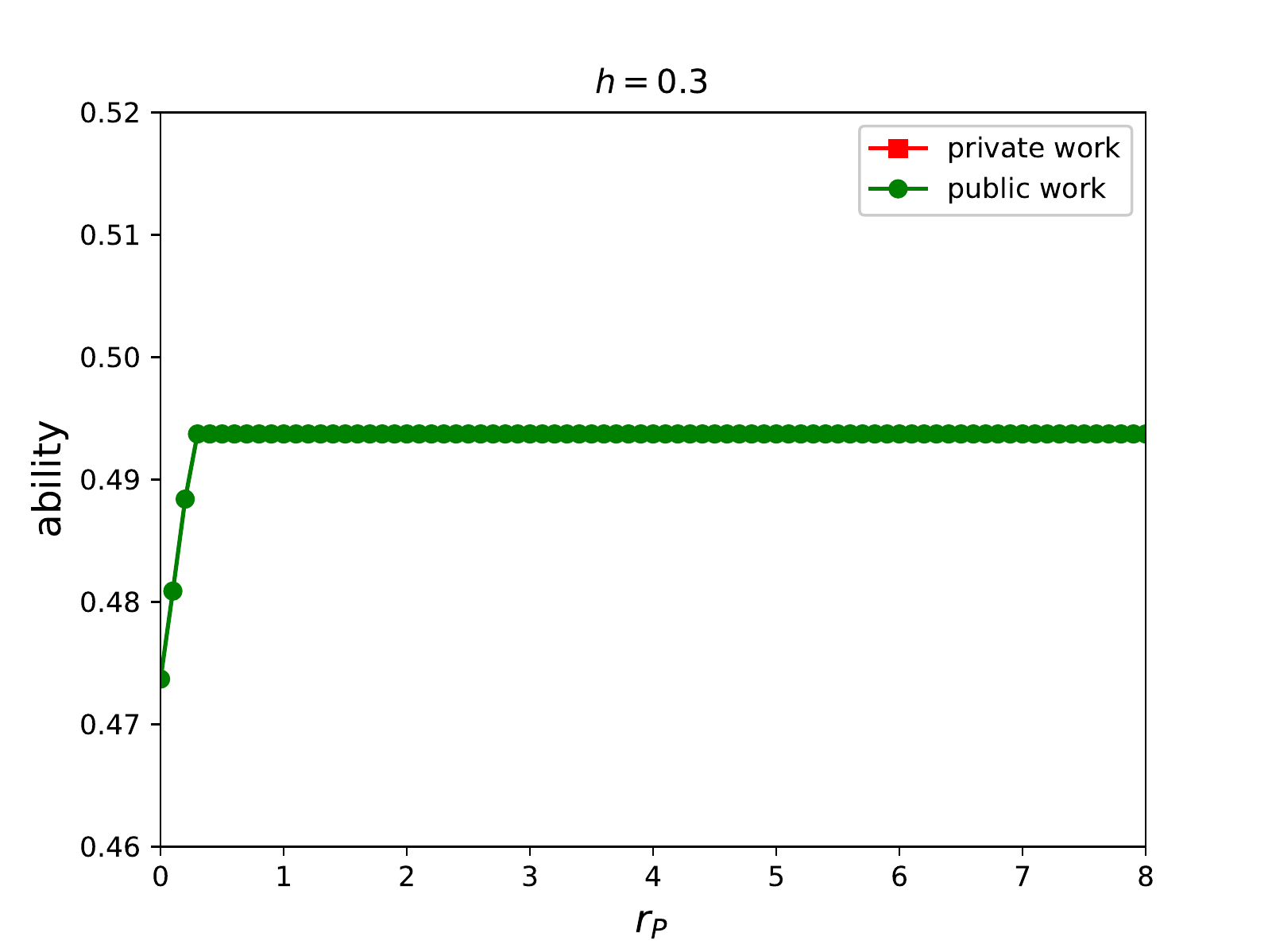}
        \caption{}
        \label{fig:fig5_3}
    \end{subfigure}
    \begin{subfigure}[b]{0.43\textwidth}
        \includegraphics[width=\textwidth]{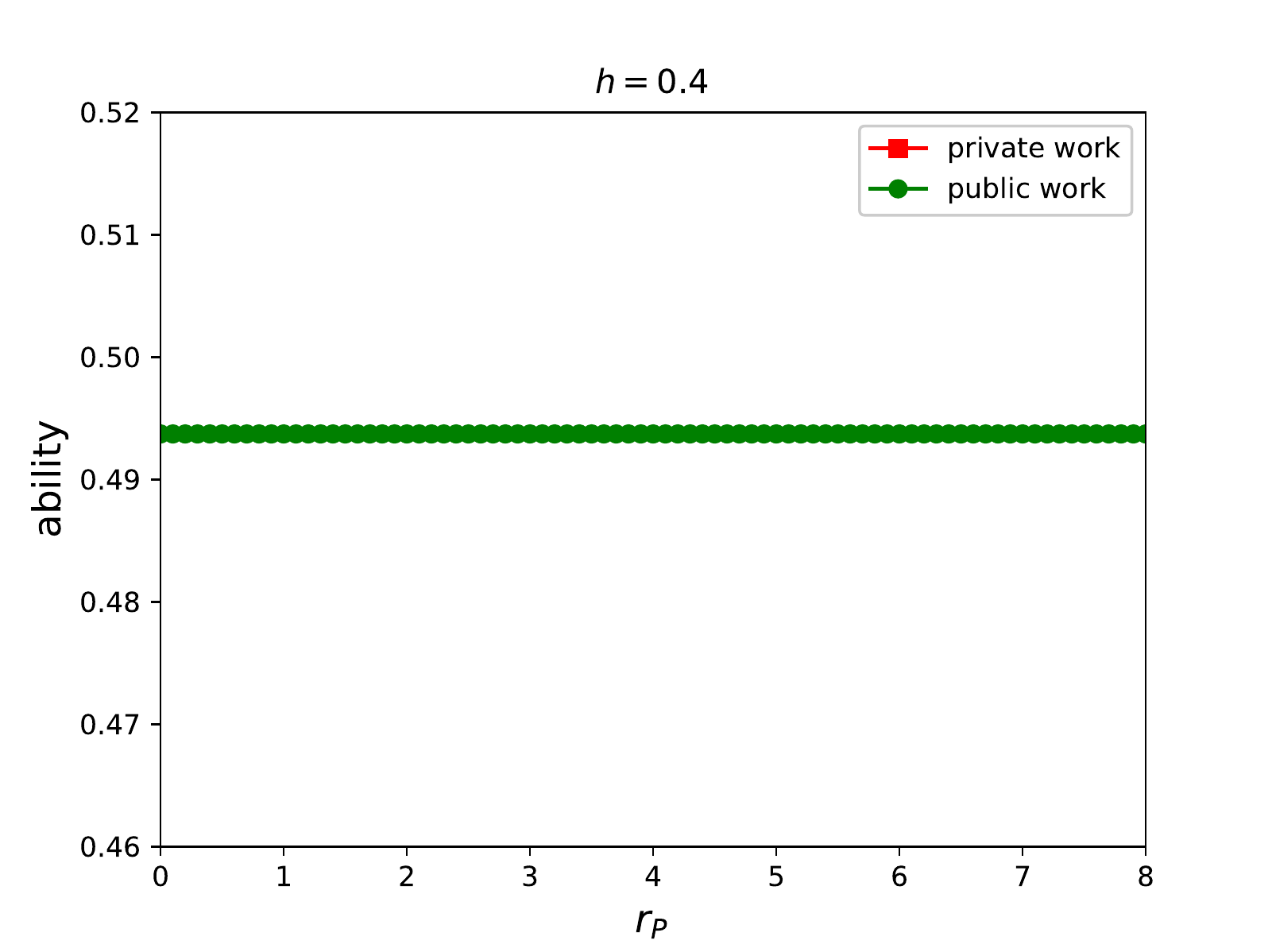}
        \caption{}
        \label{fig:fig5_4}
    \end{subfigure}
    \caption{Work ability as a function of $r_P$ for regular and public work with $h$ = 0, 0.1, 0.3 and 0.4.}
    \label{fig:fig5}
\end{figure}

For the next experiment, we study the impact of work effort and  human capital on the social welfare, represented by the system output. Figure \ref{fig:fig4} shows that the effort of type-$R$ work is higher than that of type-$P$ work near the phase transition point. For type-$R$ work, the level of effort almost stays unchanged, which is $e_i \approx 0.6$ for $s_i = R$. However, the effort of type-$P$ work increases linearly with $r_P$. The slope of the effort curve is calculated as 0.0075, which can be observed from Figures \ref{fig:fig4_1} to  \ref{fig:fig4_4}. When $h = 0$, the phase transition occurs at $r_P = 5.2$. The work efforts are $e_i = 0.585$ and $e_i = 0.54$ for $s_i = R$ and $s_i =P$ at the phase transition point, respectively. Similarly, the phase transition occurs at near $r_P = 3.7$ for $h = 0.1$ and the work efforts are $e_i = 0.595$ and $e_i = 0.53$ for $s_i = R$ and $s_i =P$, respectively. When $h = 0.3$, the phase transition occurs at an much earlier point of $r_P=0.1$ and the level of effort is slightly higher than 0.50 for public work. At last, when the subsidy is as high as 0.4, all participants leave the type-$R$ work. 

So far, we have seen that people in type-$R$ work tend to engage more effort than people in type-$P$ work. The level of effort of type-$P$ work only increases linearly with the  employee productivity in the public sector, and they never reach the same level of the private sector.

Furthermore, we examine the relationship between human capital $\gamma_i$ and employee productivity $r_P$. The results are illustrated in Figure  \ref{fig:fig5}. When phase transition occurs, people with lower human capital remains in the public sector, and people with higher human capital choose private work. Averagely speaking, the human capital of type-$R$ work is higher than that of type-$P$ work.

\begin{figure}[htb]
    \centering
    \begin{subfigure}[b]{0.43\textwidth}
        \includegraphics[width=\textwidth]{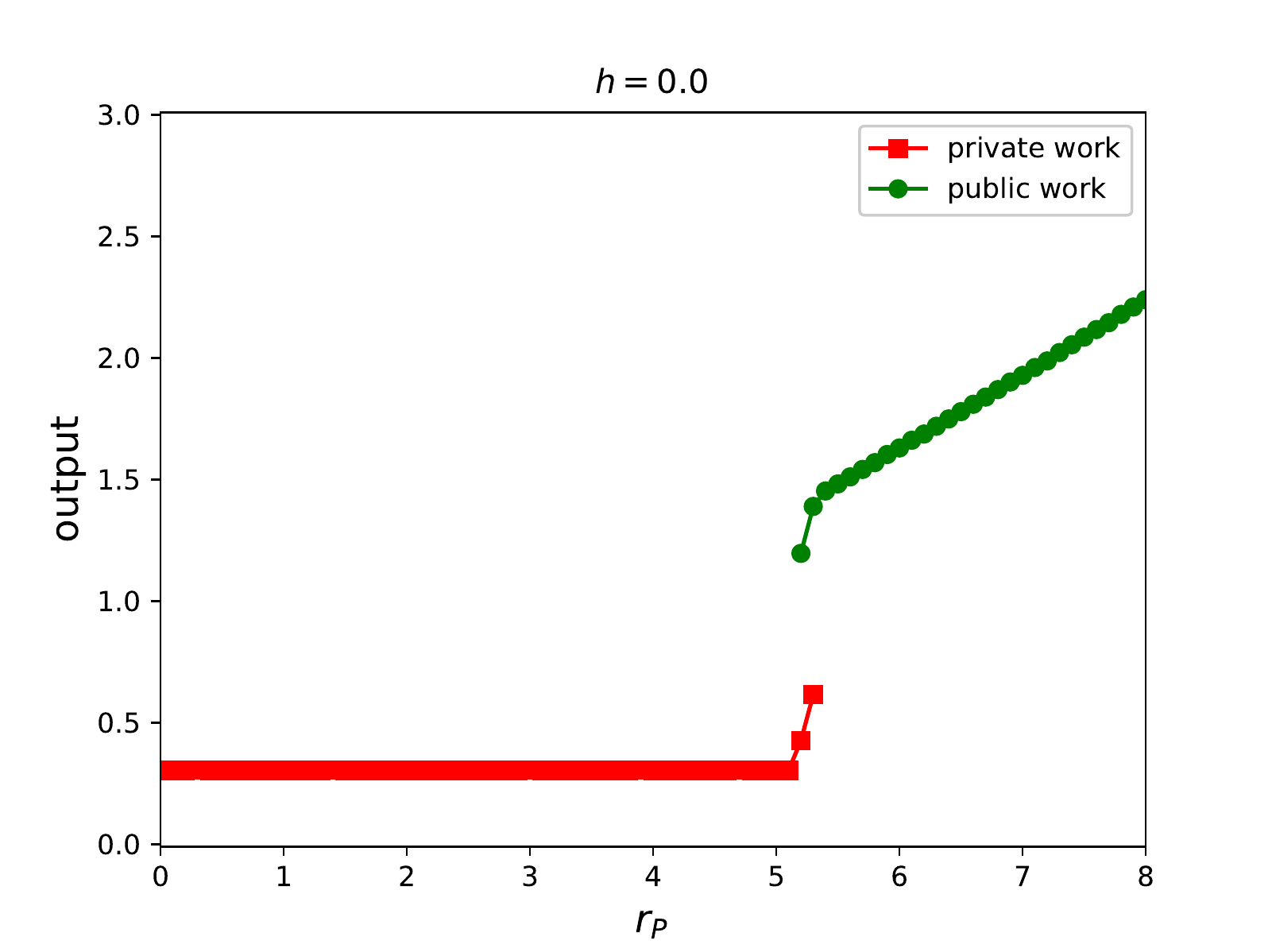}
        \caption{}
        \label{fig:fig6_1}
    \end{subfigure}
    \begin{subfigure}[b]{0.43\textwidth}
        \includegraphics[width=\textwidth]{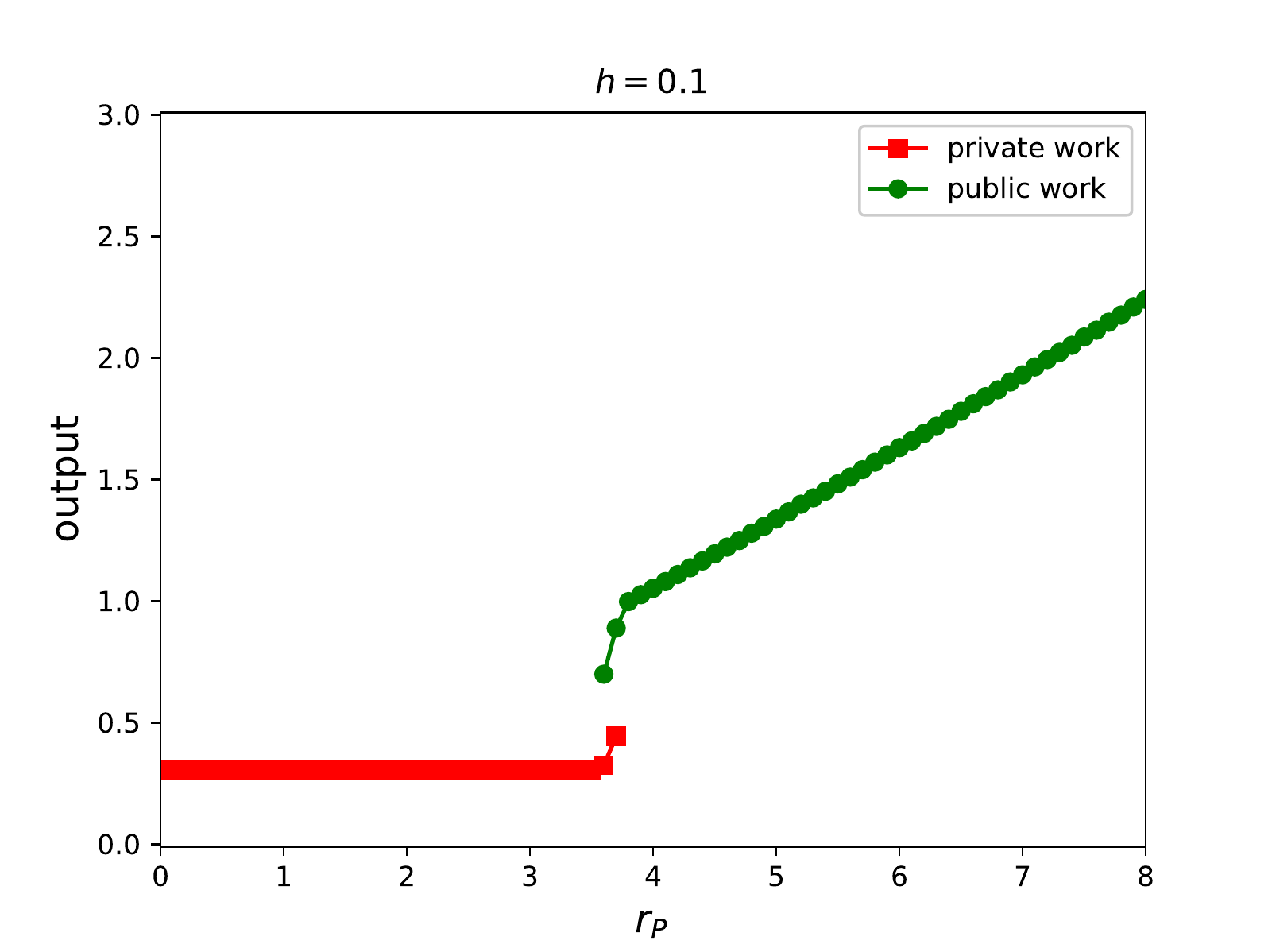}
        \caption{}
        \label{fig:fig6_2}
    \end{subfigure}
    \begin{subfigure}[b]{0.43\textwidth}
        \includegraphics[width=\textwidth]{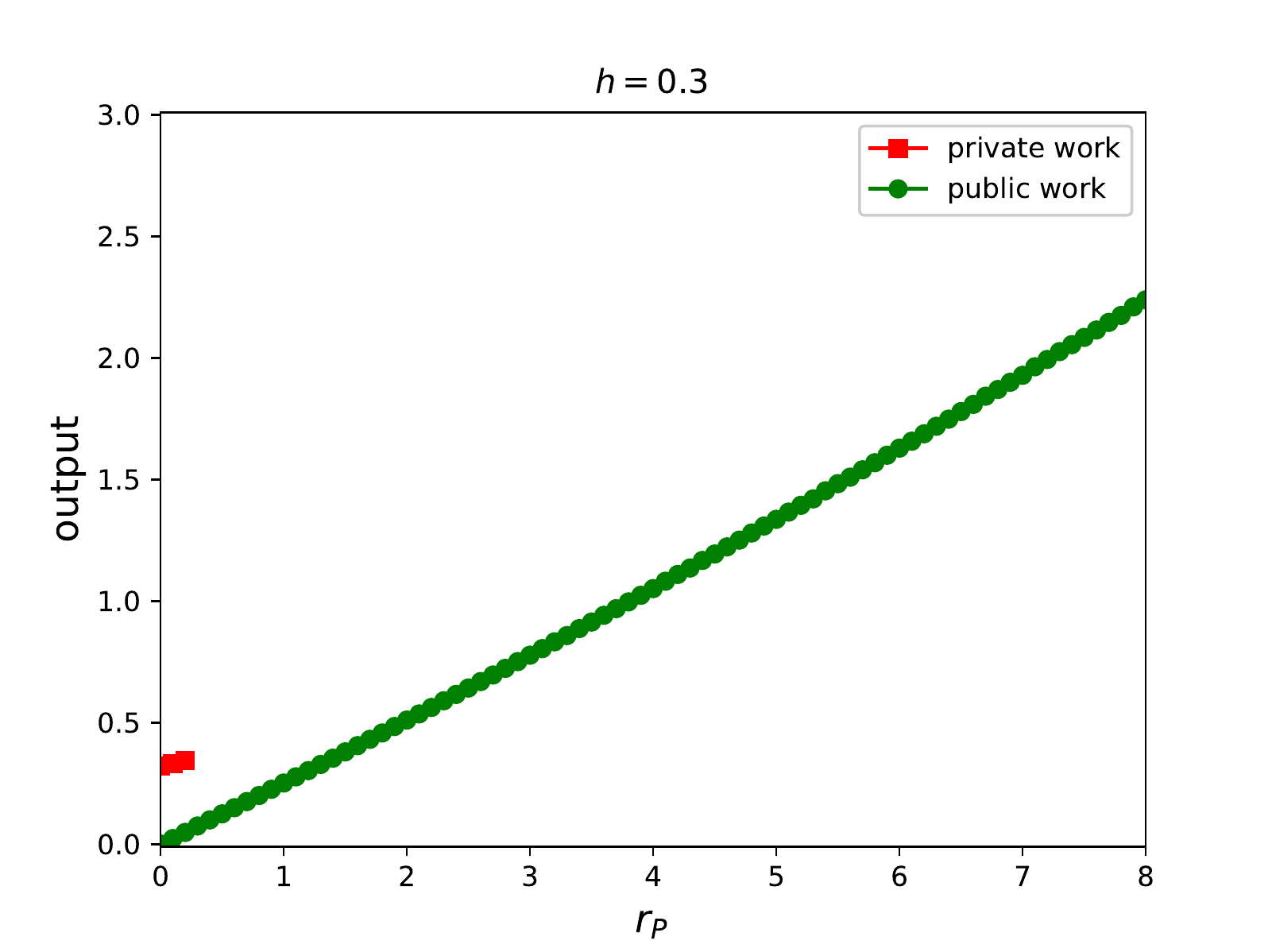}
        \caption{}
        \label{fig:fig6_3}
    \end{subfigure}
    \begin{subfigure}[b]{0.43\textwidth}
        \includegraphics[width=\textwidth]{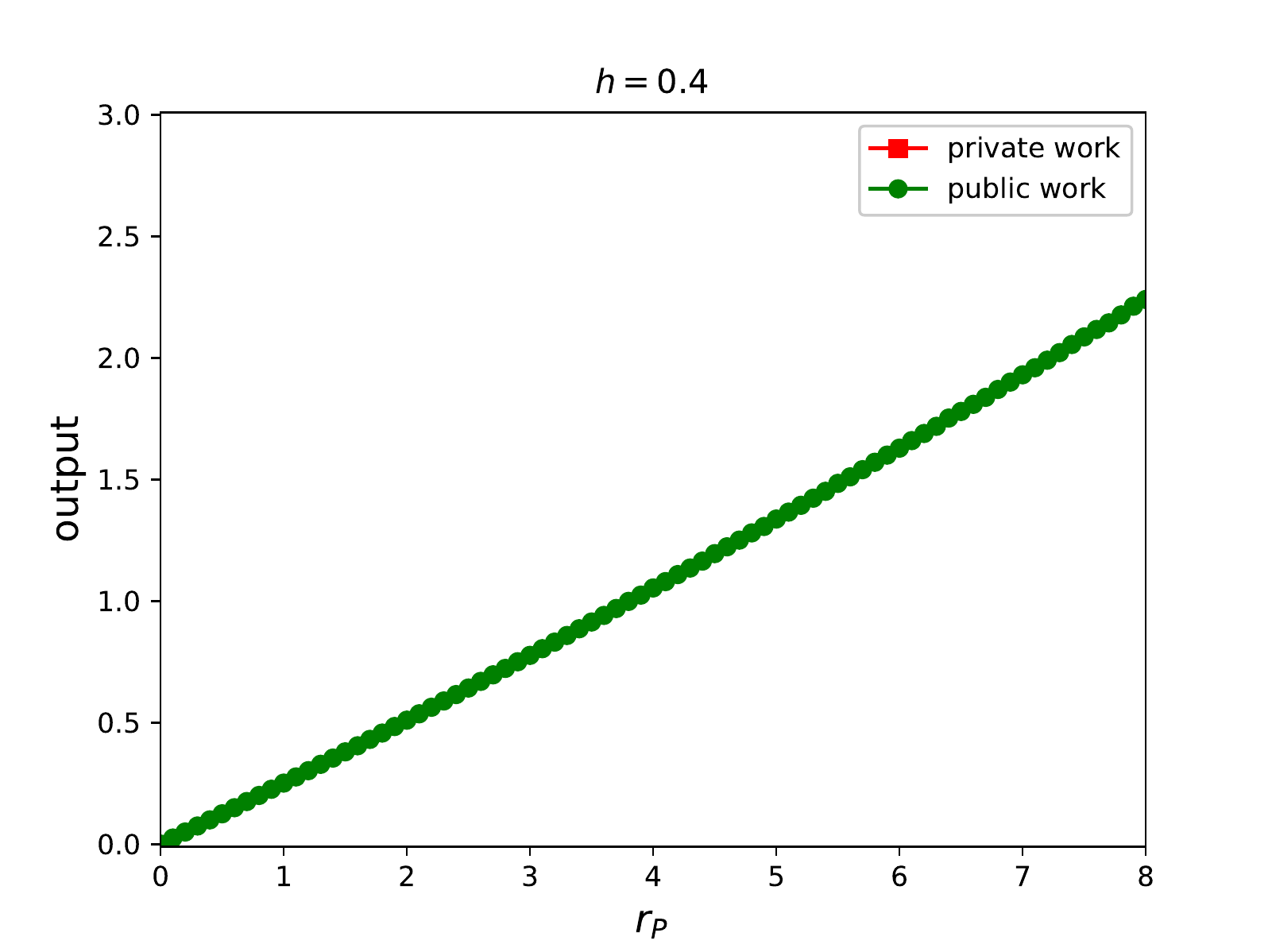}
        \caption{}
        \label{fig:fig6_4}
    \end{subfigure}
    \caption{Work output as a function of $r_P$ for public and private work with $h$ = 0, 0.1,0.3 and 0.4.}
    \label{fig:fig6}
\end{figure}

To see how the unsubsidized output of our system changes with employee productivity, we summarize the previous experiments from another perspective, as correspondingly shown by pairs of Figures \ref{fig:fig5_1} with \ref{fig:fig6_1},  Figures \ref{fig:fig5_2} with \ref{fig:fig6_2}, Figures \ref{fig:fig5_3} with \ref{fig:fig6_3}, and Figures \ref{fig:fig5_4} with \ref{fig:fig6_4}. Even if people in the private sector possess higher human capital and work harder, as evidenced by Figures \ref{fig:fig5_1} and \ref{fig:fig5_2}, the system outputs of the public sector are significantly higher under the circumstances of no or low subsidy, as evidenced by Figures \ref{fig:fig6_1} and \ref{fig:fig6_2}. The net output of type-$R$ work is mostly insensitive to the changes of $r_P$, {\em i.e.}, $o \approx 0.3$.  
For type-$P$ work, its net output increases proportionally with $r_P$ at a slope of 0.28. However, when the subsidy is very high ($h = 0.4$), as shown by Figure \ref{fig:fig6_4}, the phase transition happens at a point of zero employee productivity in public section ($r_P = 0$). The conclusion follows naturally: public work can lead to a higher social welfare than private work provided that subsidy is low.

\subsection*{(4) Subsidy and social welfare}
\begin{figure}[tb]
    \centering
    \begin{subfigure}[b]{0.43\textwidth}
        \includegraphics[width=\textwidth]{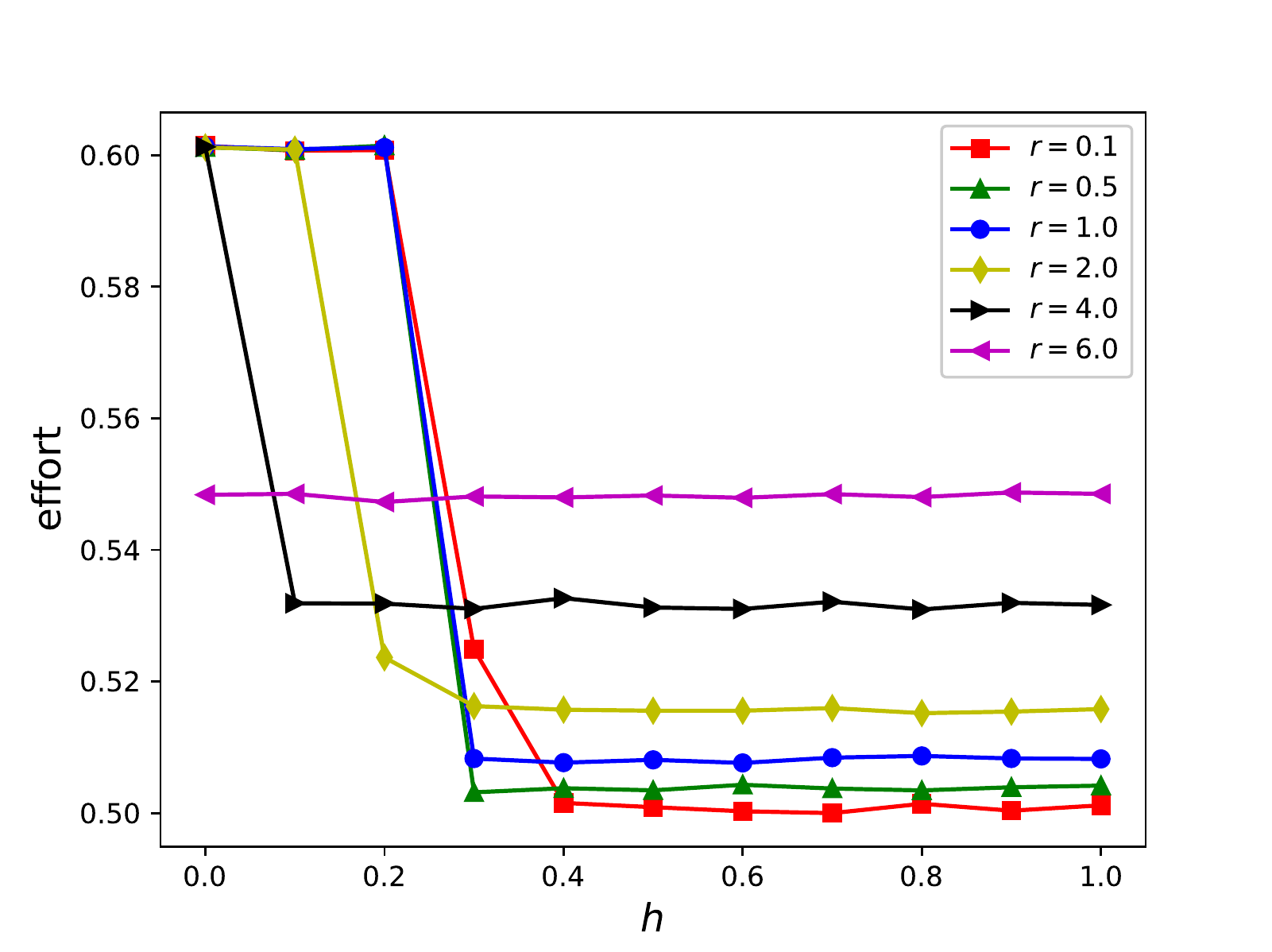}
        \caption{}
        \label{fig:fig7_1}
    \end{subfigure}
    \begin{subfigure}[b]{0.43\textwidth}
        \includegraphics[width=\textwidth]{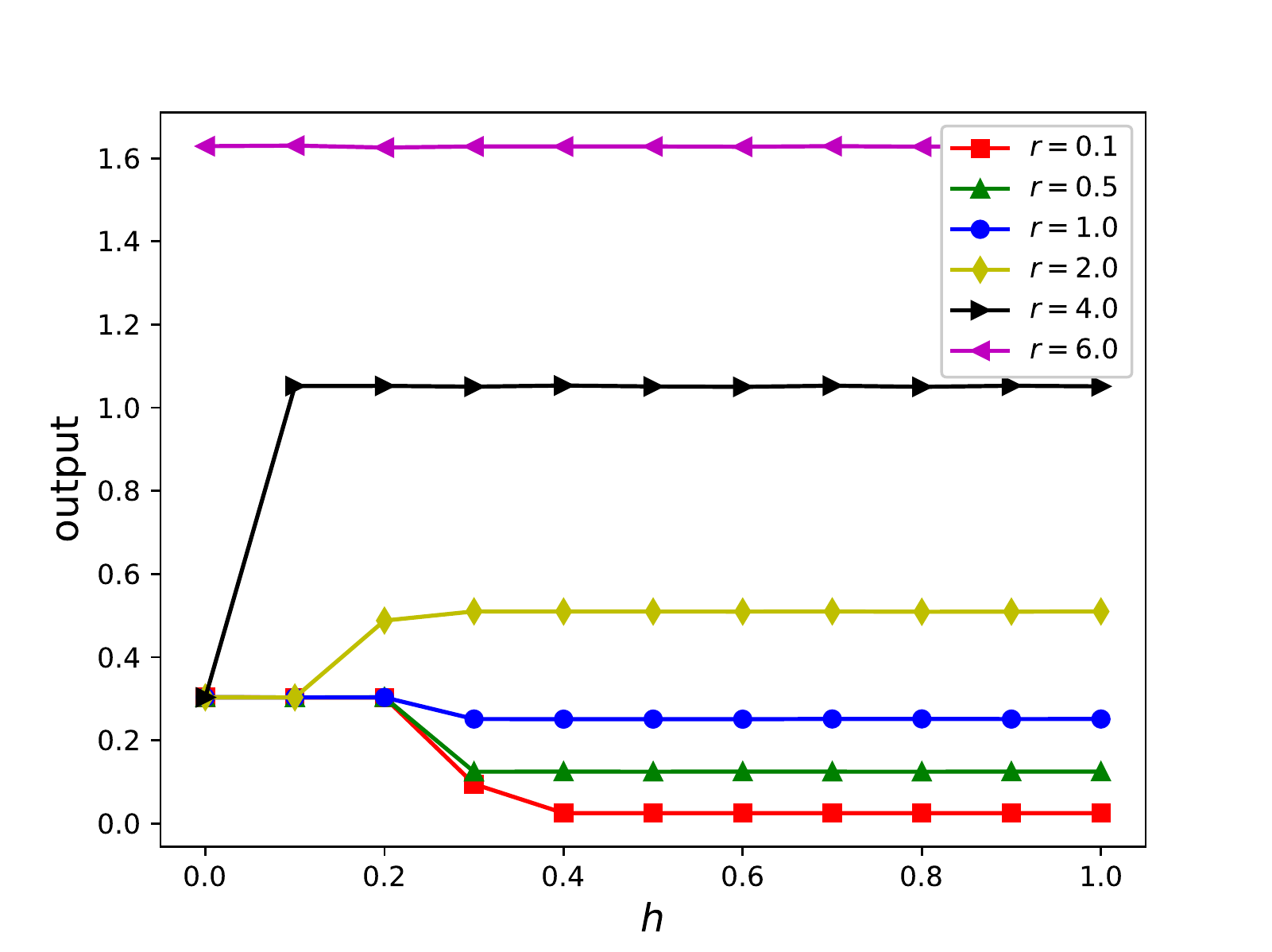}
        \caption{}
        \label{fig:fig7_2}
    \end{subfigure}
    \caption{ (1) Effort in the system as a function of $h$ with $r_P$ = 0.1, 0.5, 1.0, 2.0, 4.0 and 6.0. (2) System output as a function of $h$ with $r$ = 0.1, 0.5, 1.0, 2.0, 4.0 and 6.0.}
    \label{fig:fig7}
\end{figure}

For our final experiment, we examine the average effort of all participants in the system and the relationship between the system output and external subsidy. Figure \ref{fig:fig7_1} shows that the average effort in the entire system decreases with the amount of external subsidy. The only exception is that, when $r_P$ is very high, {\em e.g.},  $r_P = 6$, the average effort becomes flat. For $r_P =$4, 2, and 1, when subsidy increases, the average effort first decreases and then stays at a level of 0.53, 0.515, and 0.51, respectively.  Figure \ref{fig:fig7_2} plots the relationship between external subsidy and net system output, or namely social welfare. When employee productivity in the public sector is relatively large, {\em e.g.}, $r_P$ = 2, 4 or 6, adding subsidy can increase the net output of the system. For example, when $r_P = 2$, as the subsidy value increases from 0.1 to 0.2, the net output increases from 0.3 to 0.5. However, when employee productivity in the public sector is low, {\em e.g.}, $r_P$ = 0.1, 0.5, or 1, adding subsidy can be detrimental to social welfare. For example, when $r_P = 0.5$, as the subsidy increases from 0.3 to 0.4, the net output drops from 0.3 to 0.1.

Generally speaking, increasing external subsidy can encourage more participants to choose public work. However, since work effort of the public sector is lower than that of the private sector, high level of subsidy will reduce the overall outcome. On the contrary, the system output of public work is higher than that of private work when $r_P > r_R$, and adding subsidy will increase the net output of the overall system.

\section{Concluding Remarks} \label{section:conclusion}
In this paper, we studied the dynamics of people's career choice from a perspective of complex system by first extending and then applying the spatial evolutionary public goods game model. Based on a  social value orientation implication, and for the sake of simplicity, we classified career choices into two distinctive categories: public work and private work. We also include human capital, employee productivity and personal effort to examine work output of both individuals and the system under the impact of different career choices. To make our model even more practical, we have also included subsidy effect to incentive people working in the public sector. Major results are found as followed:
\begin{enumerate}[(1)]
	\item Employee productivity of public work affects the phase transition point of individual career choices in the absence of external subsidy. When $r_P$ is low, participants gradually migrate to private work. When $r_P$ reaches the vicinity of the phase transition point, the fraction of public workers increases rapidly until all players in the system resort to public work. 	
	\item Individuals engaged in private work on average have higher levels of work effort and human capital than individuals in public work, but public work can lead to higher social output given that the level of external subsidy is low. The level of effort of public workers only increases linearly with their employee productivity. 
	\item Level of external subsidies directly affects individual career choices and system output. Providing subsidies to public work can attract more workers, even though they tend to be less productive. However, high level of subsidies for public work can also reduce individual's efforts. For public work, under the conditions of higher employee productivity, more subsidies can increase overall system output; under the condition of low employee productivity, higher level of subsidies will in turn reduce the overall output.   
\end{enumerate}

The above findings via our SEPGG model suggest that employee productivity and the level of external subsidies for public work are crucial factors to observe individual career choice. To our best knowledge, this study is the first attempt in modelling the dynamics of career choice as an evolutionary game. As we will be overwhelmingly seeing the mixed outcome of the game in real life, this study provides insights to understanding individual career choices from the perspective of complex system. The preliminary results obtained in this and future research can serve as a framework of reference for delicate policy-makings that are pertinent to setting subsidy levels for public causes and improving productivity of public workers.

\section*{Acknowledgment}\label{section:acknowledgment}
The authors are grateful to Lu Wang for discussions about career choice.


\begin{thebibliography}{10}
\expandafter\ifx\csname url\endcsname\relax
  \def\url#1{\texttt{#1}}\fi
\expandafter\ifx\csname urlprefix\endcsname\relax\def\urlprefix{URL }\fi
\expandafter\ifx\csname href\endcsname\relax
  \def\href#1#2{#2} \def\path#1{#1}\fi

\bibitem{carpenter1977career}
P.~Carpenter, B.~Foster, The career decisions of student teachers, Educational
  Research and perspectives 4~(1) (1977) 23--33.

\bibitem{beynon1998visible}
J.~Beynon, K.~Toohey, N.~Kishor, Do visible minority students of chinese and
  south asian ancestry want teaching as a career?: Perceptions of some
  secondary school students in vancouver, bc, Canadian Ethnic Studies Journal
  30~(2) (1998) 50--75.

\bibitem{career2010}
J.~Greenhaus, G.~Callanan, V.~Godshalk, Career Management (4th edition), Sage
  Publication, London, 2010.

\bibitem{aycan2003}
Z.~Aycan, S.~Fikret-Pasa, Career choices, job selection criteria, and
  leadership preferences in a transitional nation: The case of turkey, Journal
  of Career Development 30~(2) (2003) 129--144.

\bibitem{rojewski2003}
J.~W. Rojewski, H.~Kim, Career choice patterns and behavior of work-bound youth
  during early adolescence, Journal of Career Development 30~(2) (2003)
  89--108.

\bibitem{arthur2005}
M.~B. Arthur, S.~N. Khapova, C.~P. Wilderom, Career success in a boundaryless
  career world, Journal of Organizational Behavior: The International Journal
  of Industrial, Occupational and Organizational Psychology and Behavior 26~(2)
  (2005) 177--202.

\bibitem{jiang2018}
Z.~Jiang, A.~Newman, H.~Le, A.~Presbitero, C.~Zheng, Career exploration: A
  review and future research agenda, Journal of Vocational Behavior.

\bibitem{chuang2016}
A.~Chuang, C.-T. Shen, T.~A. Judge, Development of a multidimensional
  instrument of person--environment fit: The perceived person--environment fit
  scale (ppefs), Applied Psychology 65~(1) (2016) 66--98.

\bibitem{bednarska2017}
M.~A. Bednarska, Does the effect of person-environment fit on work attitudes
  vary with generations? insights from the tourism industry, International
  Journal Of Management And Economics 53~(1) (2017) 65--83.

\bibitem{van2018}
A.~E. van Vianen, Person--environment fit: A review of its basic tenets, Annual
  Review of Organizational Psychology and Organizational Behavior 5 (2018)
  75--101.

\bibitem{akosah2018systematic}
P.~Akosah-Twumasi, T.~I. Emeto, D.~Lindsay, K.~Tsey, B.~Malau-Aduli, A
  systematic review of factors that influence youths career choices--the role
  of culture, in: Frontiers in Education, Vol.~3, Frontiers, 2018, p.~58.

\bibitem{schwartz1992universals}
S.~H. Schwartz, Universals in the content and structure of values:
  {T}heoretical advances and empirical tests in 20 countries, Advances in
  Experimental Social Psychology 25 (1992) 1--65.

\bibitem{schwartz2012refining}
S.~H. Schwartz, J.~Cieciuch, M.~Vecchione, E.~Davidov, R.~Fischer,
  C.~Beierlein, A.~Ramos, M.~Verkasalo, J.-E. L{\"o}nnqvist, K.~Demirutku,
  et~al., Refining the theory of basic individual values., Journal of
  personality and social psychology 103~(4) (2012) 663.

\bibitem{de2008value}
J.~I. De~Groot, L.~Steg, Value orientations to explain beliefs related to
  environmental significant behavior: How to measure egoistic, altruistic, and
  biospheric value orientations, Environment and Behavior 40~(3) (2008)
  330--354.

\bibitem{brown2002role}
D.~Brown, The role of work and cultural values in occupational choice,
  satisfaction, and success: A theoretical statement, Journal of counseling \&
  development 80~(1) (2002) 48--56.

\bibitem{forsyth2006conflict}
D.~R. Forsyth, Conflict, Forsyth, DR, Group Dynamics (2006) 388--389.

\bibitem{bandura1963social}
A.~Bandura, R.~H. Walters, Social learning and personality development.

\bibitem{krumboltz1976}
J.~D. Krumboltz, A.~M. Mitchell, G.~B. Jones, A social learning theory of
  career selection, The counseling psychologist 6~(1) (1976) 71--81.

\bibitem{dur2018}
R.~Dur, M.~van Lent, Serving the public interest in several ways: Theory and
  empirics, Labour Economics 51 (2018) 13--24.

\bibitem{perc2013}
M.~Perc, J.~G{\'o}mez-Garde{\~n}es, A.~Szolnoki, L.~M. Flor{\'\i}a, Y.~Moreno,
  Evolutionary dynamics of group interactions on structured populations: a
  review, Journal of the royal society interface 10~(80) (2013) 20120997.

\bibitem{hauert2006synergy}
C.~Hauert, F.~Michor, M.~A. Nowak, M.~Doebeli, Synergy and discounting of
  cooperation in social dilemmas, Journal of theoretical biology 239~(2) (2006)
  195--202.

\bibitem{hardin1968}
G.~Hardin, The tragedy of the commons, Science 162~(3859) (1968) 1243--1248.

\bibitem{kim2015spatial}
J.~Kim, H.~Chae, S.-H. Yook, Y.~Kim, Spatial evolutionary public goods game on
  complete graph and dense complex networks, Scientific reports 5 (2015) 9381.

\bibitem{lahkar2019evolutionary}
R.~Lahkar, S.~Mukherjee, Evolutionary implementation in a public goods game,
  Journal of Economic Theory.

\bibitem{nowak1992}
M.~A. Nowak, R.~M. May, Evolutionary games and spatial chaos, Nature 359~(6398)
  (1992) 826.

\bibitem{szabo1998}
G.~Szab{\'o}, C.~T{\H{o}}ke, Evolutionary prisoner’s dilemma game on a square
  lattice, Physical Review E 58~(1) (1998) 69.

\bibitem{szabo2007}
G.~Szab{\'o}, G.~Fath, Evolutionary games on graphs, Physics reports 446~(4-6)
  (2007) 97--216.

\bibitem{roca2009}
C.~P. Roca, J.~A. Cuesta, A.~S{\'a}nchez, Evolutionary game theory: Temporal
  and spatial effects beyond replicator dynamics, Physics of life reviews 6~(4)
  (2009) 208--249.

\bibitem{perc2010}
M.~Perc, A.~Szolnoki, Coevolutionary games—a mini review, BioSystems 99~(2)
  (2010) 109--125.

\bibitem{perc2017}
M.~Perc, J.~J. Jordan, D.~G. Rand, Z.~Wang, S.~Boccaletti, A.~Szolnoki,
  Statistical physics of human cooperation, Physics Reports 687 (2017) 1--51.

\bibitem{kollock1998}
P.~Kollock, Social dilemmas: The anatomy of cooperation, Review of Sociology
  24~(1) (1998) 183--214.

\bibitem{szolnoki2011}
A.~Szolnoki, G.~Szab{\'o}, M.~Perc, Phase diagrams for the spatial public goods
  game with pool punishment, Physical Review E 83~(3) (2011) 036101.

\bibitem{helbing2010}
D.~Helbing, A.~Szolnoki, M.~Perc, G.~Szab{\'o}, Punish, but not too hard: how
  costly punishment spreads in the spatial public goods game, New Journal of
  Physics 12~(8) (2010) 083005.

\bibitem{shi2010}
D.-M. Shi, Y.~Zhuang, B.-H. Wang, Group diversity promotes cooperation in the
  spatial public goods game, EPL (Europhysics Letters) 90~(5) (2010) 58003.

\bibitem{zhu2014}
C.-j. Zhu, S.-w. Sun, L.~Wang, S.~Ding, J.~Wang, C.-y. Xia, Promotion of
  cooperation due to diversity of players in the spatial public goods game with
  increasing neighborhood size, Physica A: Statistical Mechanics and its
  Applications 406 (2014) 145--154.

\bibitem{huang2015}
K.~Huang, T.~Wang, Y.~Cheng, X.~Zheng, Effect of heterogeneous investments on
  the evolution of cooperation in spatial public goods game, PloS one 10~(3)
  (2015) e0120317.

\bibitem{hauert2002}
C.~Hauert, S.~De~Monte, J.~Hofbauer, K.~Sigmund, Volunteering as red queen
  mechanism for cooperation in public goods games, Science 296~(5570) (2002)
  1129--1132.

\bibitem{semmann2003}
D.~Semmann, H.-J. Krambeck, M.~Milinski, Volunteering leads to
  rock--paper--scissors dynamics in a public goods game, Nature 425~(6956)
  (2003) 390.

\bibitem{milinski2002}
M.~Milinski, D.~Semmann, H.-J. Krambeck, Reputation helps solve the ‘tragedy
  of the commons’, Nature 415~(6870) (2002) 424.

\bibitem{xia2016}
C.~Xia, S.~Ding, C.~Wang, J.~Wang, Z.~Chen, Risk analysis and enhancement of
  cooperation yielded by the individual reputation in the spatial public goods
  game, IEEE Systems Journal 11~(3) (2016) 1516--1525.

\bibitem{anzola2017}
D.~Anzola, P.~Barbrook-Johnson, J.~I. Cano, Self-organization and social
  science, Computational and Mathematical Organization Theory 23~(2) (2017)
  221--257.

\bibitem{sigmund2010}
K.~Sigmund, H.~De~Silva, A.~Traulsen, C.~Hauert, Social learning promotes
  institutions for governing the commons, Nature 466~(7308) (2010) 861.

\bibitem{szabo2002}
G.~Szab{\'o}, C.~Hauert, Phase transitions and volunteering in spatial public
  goods games, Physical review letters 89~(11) (2002) 118101.

\end{thebibliography}
\end{document}